\documentclass[conference]{IEEEtran}
\IEEEoverridecommandlockouts
\usepackage{cite}
\usepackage{amsmath,amssymb,amsfonts}
\usepackage{algorithmic}
\usepackage{graphicx}
\usepackage{textcomp}
\usepackage{xcolor}
\usepackage{float}
\def\BibTeX{{\rm B\kern-.05em{\sc i\kern-.025em b}\kern-.08em
    T\kern-.1667em\lower.7ex\hbox{E}\kern-.125emX}}
\begin{document}

\title{An Efficient Online Prediction of Host Workloads Using Pruned GRU Neural Nets} 

\author{\IEEEauthorblockN{Amin Setayesh}
\IEEEauthorblockA{
University of Tehran\\
amin.setayesh@ut.ac.ir}
\and
\IEEEauthorblockN{Hamid Hadian}
\IEEEauthorblockA{University of Virginia Tech\\
hamidhadian@vt.edu }
\and
\IEEEauthorblockN{Radu Prodan}
\IEEEauthorblockA{University of Klagenfurt\\
radu.prodan@aau.at}
}

\maketitle

\begin{abstract}
Host load prediction is essential for dynamic resource scaling and job scheduling in a cloud computing environment. In this context, workload prediction is challenging because of several issues. First, it must be accurate to enable precise scheduling decisions. Second, it must be fast to schedule at the right time. Third, a model must be able to account for new patterns of workloads so it can perform well on the latest and old patterns. Not being able to make an accurate and fast prediction or the inability to predict new usage patterns can result in severe outcomes such as service level agreement (SLA) misses. Our research trains a fast model with the ability of online adaptation based on the gated recurrent unit (GRU) to mitigate the mentioned issues. We use a multivariate approach using several features, such as memory usage, CPU usage, disk I/O usage, and disk space, to perform the predictions accurately. Moreover, we predict multiple steps ahead, which is essential for making scheduling decisions in advance. Furthermore, we use two pruning methods: L1 norm and random, to produce a sparse model for faster forecasts. Finally, online learning is used to create a model that can adapt over time to new workload patterns.
\end{abstract}


\section{Introduction}
Cloud computing has gained extreme usage and popularity in recent years, and the global market size reached an estimation of 482 billion dollars in 2022 \cite{fact.mr}. Many end-users and IT companies use cloud computing services such as storage, hosting, computing, and software services due to their financial efficiency gained by the elastic resource provisioning and pay-as-you-go plans \cite{vaquero2008break}.
Cloud environments and clusters provide on-demand access over a shared pool of computing resources for for IoT applications and scientific applications and host a vast number of applications with various resource demands and optimization objectives\cite{song2018host}\cite{validi2022hybrid}. The dynamic workload of applications and the ever-changing need of users make it hard to estimate resource utilization accurately, which may result in failure or a reduction in Quality of service (QoS) due to inaccurate resource allocation decisions.

Optimal allocation of resources ensures satisfying users' needs and minimizing cost. Keeping all the computational resources active results in over-provisioning and is not efficient. On the other hand, provisioning insufficient resources causes violations of the SLA~\cite{gupta2017online},~\cite{kaur2015energy},~\cite{farrokh2022sp}. The approach towards resource provisioning could be either reactive or proactive.
In the reactive method, the application constantly monitors some metrics, like (CPU usage and waiting time in the queue) and allocates resources accordingly. Still, the main problem is that the reactive method may perform late provisions because of identifying insufficient resources at the exact time they happen. However, the proactive approach tries to anticipate the required resources several time steps ahead to deal with lacking resources in advance~\cite{kaur2015energy},~\cite{messias2016combining}. Acquiring proper insight into future proactive resource utilization in cloud environments can pave the way to have a better scheduling plan to spread tasks across virtual machines. in addition, an accurate estimation helps applications to exploit auto-scaling and migration techniques before encountering a heavy workload.

Resource prediction mathematically can be modeled by time series. In time series prediction, previous trends of data are used to predict new events in the future~\cite{gupta2017resource}. Several statistical methods, like regressive integrated moving average (ARIMA)~\cite{sudhakar2018workload} and Markov models~\cite{surya2021prediction},~\cite{hadian2023elastic} are commonly used to forecast future workloads. As workload trends are non-linear, AI-based methods like neural networks, support vector machines (SVM), and reinforcement learning (RL) for workload prediction are also utilized by researchers to minimize prediction error~\cite{gupta2017resource},~\cite{xu2022esdnn}. Due to the ability of recurrent neural networks (RNNs) to capture long-range patterns of workloads, these neural networks have recently been widely used for host load prediction~\cite{song2018host},~\cite{gupta2017resource}. Because of not being able to carry learning gradients well in deep traditional RNNs, several new variants, such as long short-term memory (LSTM) \cite{lstm}, and gated recurrent unit (GRU)~\cite{chung2014empirical}, are introduced to perform better on deep neural networks.

For the prediction of resource utilization, various parameters affect each other. For example, the amount of used memory or the number of I/O requests can directly impact CPU utilization. Multivariate models consider the effect of different features on each other and make a more accurate model to predict future trends~\cite{cheng2019gru}. In addition, being only one step ahead of time while predicting resource utilization does not guarantee precise provisioning of resources. Moreover, The workload in cloud environments is quite variant and a trained model cannot fully exhibit the trends. Hence, using adaptive models enables parameter adjustments at run time~\cite{gupta2017online},~\cite{gupta2020online}. Another vital challenge is that neural network models require considerable computation resources like memory and CPU, so pruning aims to drop some of the weights to reduce prediction time and resource consumption. Specifically, making a real-time prediction for critical systems is crucial to providing resources as soon as possible.

In this regard, the main contributions of the paper would be: 

\begin{enumerate}

\item \textbf{\emph{Analysis}}
of different GRU network hyperparameters in addition to comparison with baseline models.

\item \textbf{\emph{Development}}
of a multivariate multistep GRU network being able to predict host workloads accurately.

\item \textbf{\emph{Comparison}}
of two different Sparse GRU methods: random structured and L1 norm approaches, developed to reduce prediction time with slight accuracy decay.

\item \textbf{\emph{Examination}}
of online adaptation using two optimization methods: GD and L-BFGS, used to let our model improve over time by observing new patterns.

\end{enumerate}

The rest of the paper is organized as follows:
First, the related work is introduced in Section \ref{background}. Next, the methodology being used is described thoroughly in Section \ref{method}. Then, our experiments are presented in Section \ref{experiments}. Finally, our conclusions and future work are discussed in Section \ref{conclusion}.

\section{Related Work}
\label{background}
Various studies have investigated predicting the workload of cloud hosts and applied different methods to solve the workload prediction problem, improving models' accuracy. These methods can be classified into four broad categories: statistical, machine learning, neural network, and hybrid. We emphasize the neural network approaches more as our study falls in this category.

Using Regression methods, Gupta et al.~\cite{liu2016quantitative} provide an ensemble approach, which uses multilinear regression in combination with the moving average (MV), weighted average, and neural networks. The Ensemble Workload Prediction (EnWoP) proposed in this study uses a new prediction evaluation parameter called C-Rate to predict Google Cluster trace workloads.

Analytical models such as autocorrelation function (ACF), autoregressive integrated moving average (ARIMA) \cite{calheiros2014workload}, and seasonal integrated ARMA (SARIMA) \cite{kumar2016forecasting} are widely used for time series prediction. Many works utilize these approaches to predict host workloads. In \cite{calheiros2014workload}, the authors used the ARIMA model to predict workload. Their results show an average accuracy of up to 91\%. Additionally, Kumar et al.~\cite{kumar2016forecasting} compared several ARMA models, such as ARIMA, seasonal integrated ARMA (SARIMA), and fractionally integrated ARMA (ARFIMA). They evaluated the models using the Wikimedia CPU, RAM, and Network traces; the researchers found ARIMA the model providing the best forecast results.

Using machine learning schemes, researchers mostly used clustering, SVM, decision trees, and random forest approaches. Zhong et al.~\cite{zhong2018load} and Nikravesh et al.~\cite{nikravesh2015towards} used SVM-based techniques to predict host loads. Zhong et al.~\cite{zhong2018load} use an SVM in combination with a wavelet transform (WT) to create the WWSVM prediction model, which they state is superior to the previous naive SVM model and ARIMA model in prediction accuracy. Nikravesh et al.~\cite{nikravesh2015towards} use SVMs and MLP-based algorithms to achieve better prediction accuracy using different algorithms for different workload patterns. Furthermore, in \cite{marcus2016workload} and \cite{yang2018intelligent} decision trees were used. In addition, Cetinski et al.~\cite{cetinski2015ame} used a random forest classifier to improve prediction accuracy. They also provided confidence factors for the results for more reliability.

Neural network-based methods have attracted significant attention in recent years for forecasting trends in time series. Primarily, recurrent neural networks and the derived models, such as Long Short-Term Memory (LSTM), are widely used to predict host loads. Kumar et al.~\cite{KUMAR2018676} exploits the LSTM-RNN method to predict the workload of different servers. They compared the LSTM-RNN results with the previously used approaches and showed this model outperforms the other methods. In \cite{ouhame2021efficient}, the authors used a convolutional neural network (CNN) in combination with LSTMs to achieve higher accuracy. They evaluated the model on the Bit-brains distributed data center and showed it has a lower mean absolute error (MAE) than the ARIMA-LSTM, VAR-GRU, and VAR-MLP methods. Furthermore, Dang-Quang et al.~\cite{dang2022efficient} proposed an autoscaling framework based on the bi-directional LSTM (Bi-LSTM) method. They used Bitbrains and Materna data centers' traces to evaluate the results. The evaluation showed that the multivariate Bi-LSTM model could outperform the univariate Bi-LSTM model in addition to LSTM and CNN-LSTM models. Cheng et al.~\cite{cheng2019gru} proposed a method using a combination of GRU and the exponential smoothing method. They experimented with different multi-step CPU utilization forecasting on the Google cluster traces. They found that the error increases with step count. However, experimental results also suggested an improvement in errors compared to previous state-of-the-art models.

Trying to create a model which improves through time and can adapt itself with variations, Kumar et al.~\cite{kumar2021self} and Amiri et al.~\cite{amiri2018online} use an online learning approach. The online learning models try to capture new patterns in workloads through time, which is essential in cloud environments. Furthermore, in \cite{liu2016online}, the authors describe an online learning method to estimate ARIMA models, which can be used by the workload prediction models using analytical approaches.

In summary, current methods in workload prediction use regression, statistical, machine learning, and specially RNN-based methods to improve model accuracy. In addition, some research has been done on faster prediction and online adaptation of cloud workloads, which is essential for cloud environments. However, to the authors' knowledge, limited work has been done to develop a fast, accurate, and capable of online adaptation model for cloud workloads at the same time.\\
To deal with the previous limitations, we use a multivariate GRU network to improve multistep predictions. We also utilize online learning and pruning methods to improve prediction accuracy over time and enhance prediction time to create a fast and adaptable model in addition to being accurate.

\section{Methodology}
\label{method}
In this section, we describe our approach to predicting host workloads and introduce the improvements made to develop a faster and more accurate model with the ability of online learning over time. First, we define the cloud host load prediction as a time series problem. Next, we introduce our approach, which uses a GRU neural network. Then, our pruning methods are introduced, enabling us to create a model predicting loads in a shorter time, which is essential in high-performance computing environments. Finally, we present our online learning method, which helps model adaptation over time.
\subsection{Workload Prediction Problem}
The host load prediction problem can be modeled by a time series prediction problem. The host loads from the past are available to this model, and the goal is to predict the future workloads based on the previous observations accurately. The simplest form of this prediction is just to predict one step ahead using the previous trends; however, it might be needed to know multi-steps ahead, and the model must also enable these kinds of predictions.

We define $F_t$ as the feature vector, representing the features available at time $t$. $F_t$ can be shown as $(F_{t}^{1}, F_{t}^{2}, F_{t}^{3}, ..., F_{t}^{N})$, where $F_{t}^{n}$ shows the value of feature $n$ at time $t$. Furthermore, we define the complete observations as $F_1, F_2, F_3, ..., F_T$.

It should be noted that the features are normalized in a range between zero and one. Let us define function $f$ as the normalizing function and the $f^{-1}$ as the reverse normalizing function. The function $f$ is applied to the feature vector $F_t$ to create the normalized feature vector, which we define as $F'_t$.

We define the single-step workload prediction problem as forecasting $F'^{i}_{t}$ from the previous $k$ workload observations, which can be shown as $F'_{t-k}, F'_{t-k+1}, ..., F'_{t-1}$. We define this predicted normalized workload as $\hat{F'^{i}_t}$, and the non-normalized predicted workload $\hat{F^{i}_t}$ is calculated by applying the reverse normalizing function $f'$. Similarly, the multi-step ahead prediction is formulated as forecasting $F'^{i}_{t}, F'^{i}_{t+1}, ..., F'^{i}_{t+m}$ from the previous $k$ workload observations, which can be shown as $F'_{t-k}, F'_{t-k+1}, ..., F'_{t-1}$. In most usages, the single-step prediction is not enough, and the multi-step forecast is used for better knowledge of future workloads.
\subsection{GRU}
Gated recurrent neural networks (GRU) were introduced by \cite{cho2014properties} as a variation of the vanilla RNNs. LSTMs were first developed to help models capture long-term dependencies, which vanilla RNNs could not previously detect due to the gradient vanishing problem. Following the LSTM method, the GRU approach was introduced, and the comparison in \cite{chung2014empirical} showed the GRU method could outperform the LSTM approach in terms of convergence and generalization on some datasets.

\begin{figure}
    \centering
    \includegraphics{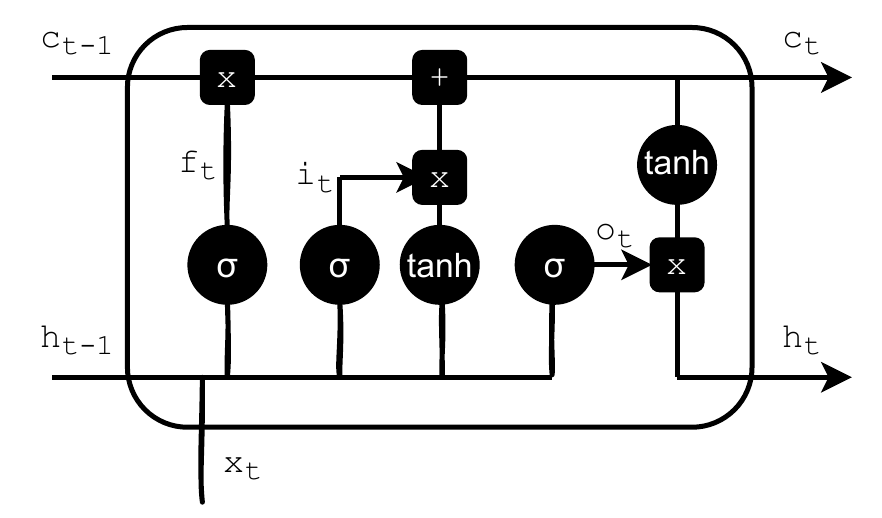}
    \caption{The architecture of an LSTM cell}
    \label{fig:LSTM}
\end{figure}

Due to the similarity of the GRU cell and the LSTM cell, we first explain the LSTM cell and then go on to explain the GRU cell. The LSTM cell uses a notion of memory, which enables the neural network to capture long-term dependencies by using the memory. LSTM unit can choose to either use the memory output or the current output. The LSTM cell is depicted in Figure \ref{fig:LSTM}.
$f_t$ is the extent to which the memory is forgotten, and it's defined as:
\begin{equation}
f_t = \sigma(W_f x_t + U_f h_{t-1} + V_f c_{t-1})
\end{equation}
$i_t$ is the extent to which the new memory is added to the previous memory values, and it's defined as:
\begin{equation}
i_t = \sigma(W_i x_t + U_i h_{t-1} + V_i c_{t-1})
\end{equation}
$V_f$ and $V_i$ are diagonal matrices \cite{chung2014empirical}.

The memory is then updated using $f_t$ and $i_t$ as the rate the memory is going to be forgotten and the rate the new memory is learned, respectively. final memory $c_t$ is calculated as:
\begin{equation}
c_t = f_t c_{t-1} + i_t \Tilde{c_t}
\end{equation}
Where $\Tilde{c_t}$ is the new memory content, and it's defined as:
\begin{equation}
    \Tilde{c_t} = tanh(W_c x_t + U_c h_{t-1})
\end{equation}
$h_t$ is the activation, and it's calculated as:
\begin{equation}
h_t = o_t tanh(c_t)
\end{equation}
where $o_t$ is the output gate.

\begin{figure}
    \centering
    \includegraphics{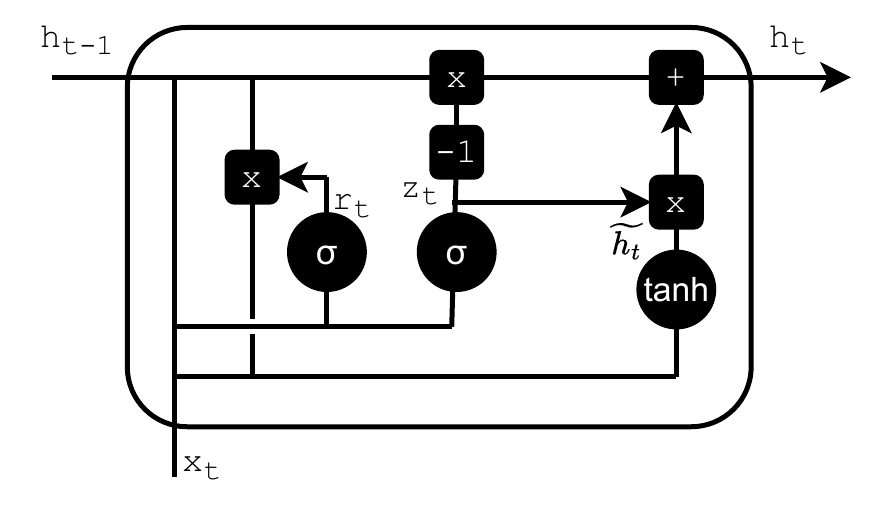}
    \caption{The architecture of a GRU cell}
    \label{fig:GRU}
\end{figure}

Unlike the LSTM cell, the GRU cell doesn't have a separate memory cell. Figure \ref{fig:GRU} shows the illustration of a GRU cell.
$h_t$ shows the activation, calculated as the combination of the previous activations and the candidate activate function, which we introduce later. $h_t$ is defined as:
\begin{equation}
h_t = (1 - z_t) h_{t-1} + z_t \Tilde{h_t}
\end{equation}
where $z_t$ is the update gate, which shows how much the activation will be updated. In addition, $\Tilde{h_t}$ is the candidate activation function.\\
The update gate is defined as:
\begin{equation}
z_t = sigma(W_z x_t + U_z h_{t-1})
\end{equation}
The candidate activation function $\Tilde{h_t}$ is defined as:
\begin{equation}
\Tilde{h_t} = tanh(W x_t + U(r_t \odot h_{t-1}))
\end{equation}
where $r_t$ is the reset gate and determines how much info will be forgotten. The reset gate is calculated as:
\begin{equation}
r_t = \sigma(W_r x_t + U_r h_{t-1})
\end{equation}
\subsection{Model Pruning}
Most neural network models keep numerous weights as their parameters after the training process. However, this can increase the model prediction time because of more computations involved. Model pruning is the process of removing some model parameters to achieve better model speed without a notable drawback to the model's accuracy. It is essential to mention that cloud computing workloads are often generated with a high ratio, and a faster model can help make better scheduling decisions in advance. Therefore, we used pruning algorithms to enhance the model's speed with minimum accuracy drop.

Structured and unstructured methods are the main approaches used to achieve model pruning. Unstructured pruning is the process of setting zero values for the individual model's weights. On the other hand, structured pruning removes entire layers or neurons to achieve a lighter model. Most current deep learning frameworks just set zero values for model weights while performing unstructured pruning. Because of this, unstructured pruning does not result in notable speed changes since the zero values are multiplied as before, and the total computations remain the same to some degree. We used two structured pruning methods to obtain better speed-ups: random and L1 norm structured approaches. The random structured method starts to choose model parameters randomly and then removes them iteratively. However, the L1 norm structured method removes model parameters by calculating the L1 norms and iteratively removing parameters with the lowest magnitudes. L1 norm of a vector $V$ is defined as the following:
\begin{equation}
    L1(V) = \sum_{i=1}^{N} v_i
\end{equation}
Where $N$ is the number of elements in the vector, and $v_i$ shows the $i$th element of the vector $V$.
\subsection{Online Learning}
Traditional workload prediction methods mostly perform initial training on the data, and no approach is suggested to improve the trained model while new workloads are observed. Consequently, the model may become inaccurate over time because of the inability to utilize new patterns generated in the workloads' data. Although the inaccuracy can be avoided by retraining the model on the newly developed data, the model forecasts inaccurate predictions in the interval before retraining. We use an online learning approach to overcome the limitations of the mentioned approaches.

\begin{figure}
    \centering
    \includegraphics[width=0.47\textwidth]{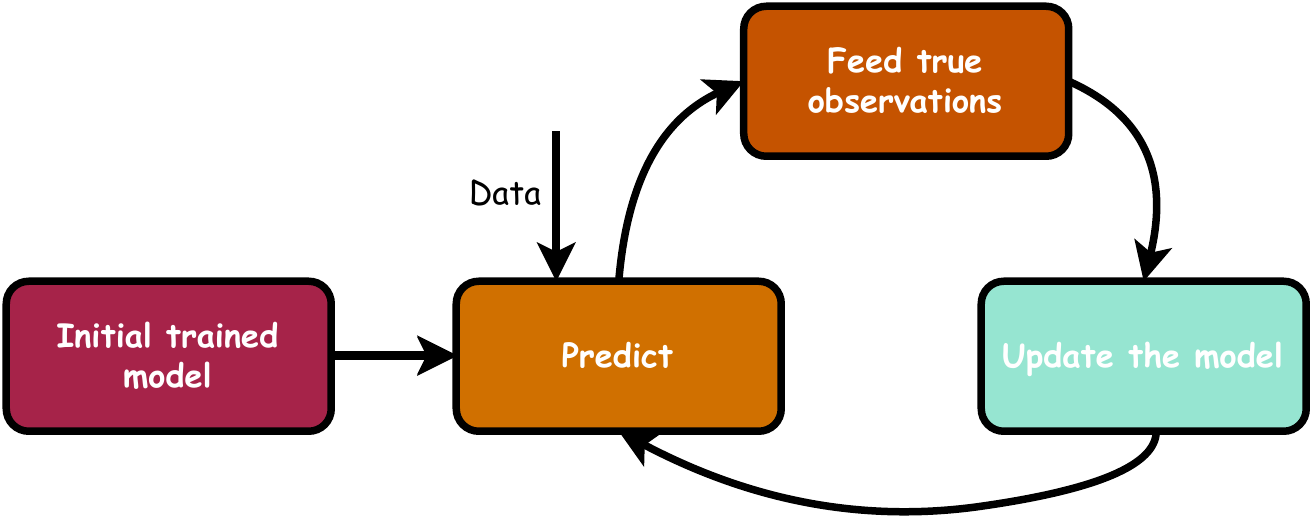}
    \caption{Overview of the online learning approach}
    \label{fig:online-learning}
\end{figure}

Online learning is the practice of extracting useful information from the newly observed data and utilizing the data to develop a more accurate model. Figure \ref{fig:online-learning} depicts an abstract overview of online learning. This method is beneficial since the model can get better over time. Moreover, This approach is especially helpful while forecasting server workloads due to the fast generation of new workloads, which may contain different usage patterns.

We utilized two optimization methods: gradient descent (GD) and Limited-memory BFGS (L-BFGS). GD is a first-order optimization approach widely used in machine learning optimization problems. GD works by updating the weights using the partial derivative of the current model error to the weight being updated. Equation \ref{eq:GD} shows the update performed in the GD method, where $\alpha$ is the learning rate and $\partial J(\Theta)$ shows the cost function calculated from the model parameters $\Theta$.
\begin{equation}
    weight_{updated} = weight_{old} - \alpha \frac{\partial J(\Theta)}{\partial weight_{old}}
\label{eq:GD}
\end{equation}
L-BFGS is a quasi-Newton family algorithm that utilizes the objective function's second-order derivatives (Hessians) while using limited memory. The following update equation is used iteratively in the BFGS algorithm to optimize the objective function:
\begin{equation}
\begin{split}
    weight_{updated} = weight_{old} - H(weight_{old})^{-1}\frac{\partial J(\Theta)}{\partial weight_{old}}
\end{split}
\end{equation}
The L-BFGS approach approximates the calculation of inverse Hessian $H(weight_{old})^{-1}$, which is computationally expensive.
\section{Experimental Results}
\label{experiments}
In this section, we provide the experimental results and their related discussions. First, we give an explanation of the dataset used in our study. Then, we provide our initial GRU model prediction results and compare the errors with other models. Next, the pruning results and the effect on the model's prediction time and error rate are shown. Finally, the results are shown after applying our online learning method.
\subsection{Dataset and Preprocessing}
We used the Google cluster traces dataset \cite{reiss2011google} for our experiments. The dataset provides detailed information about one Borg cell, which consists of 12,583 machines. The data is collected for the cluster from May 1, 2011, and 29 days of information is recorded. Each Borg cell consists of different machines, and the management system allocates work to them. Work arrives in the form of jobs, consisting of several tasks. A task is the smallest form of work, and its resource requirements are determined. The Google traces dataset contains information about machines, jobs, and tasks, such as job events, machine attributes, machine events, task constraints, task events, and resource usages. We emphasize these resource usages due to their relevance to our experiments. The resource usages are reported every five minutes in the dataset and are sampled approximately every one second. Then, the mean of the samples is calculated to show the average resource usage in the five-minute interval. The dataset contains data about several resources, such as average CPU usage rate, canonical memory usage, mean disk I/O time, and mean local disk space used. The average CPU usage rate is measured in units of CPU-core seconds per second. In addition, the canonical memory usage calculates the number of user-accessible pages, including page cache and excluding stale pages \cite{reiss2011google}. Moreover, the mean disk I/O time is measured in units of disk-time seconds per second and shows the sum across all disks in the machine. An overview of the dataset is provided in Table \ref{tab:google-traces-metrics}.

\begin{figure}[hbt]
    \centering
    \includegraphics[width=0.47\textwidth]{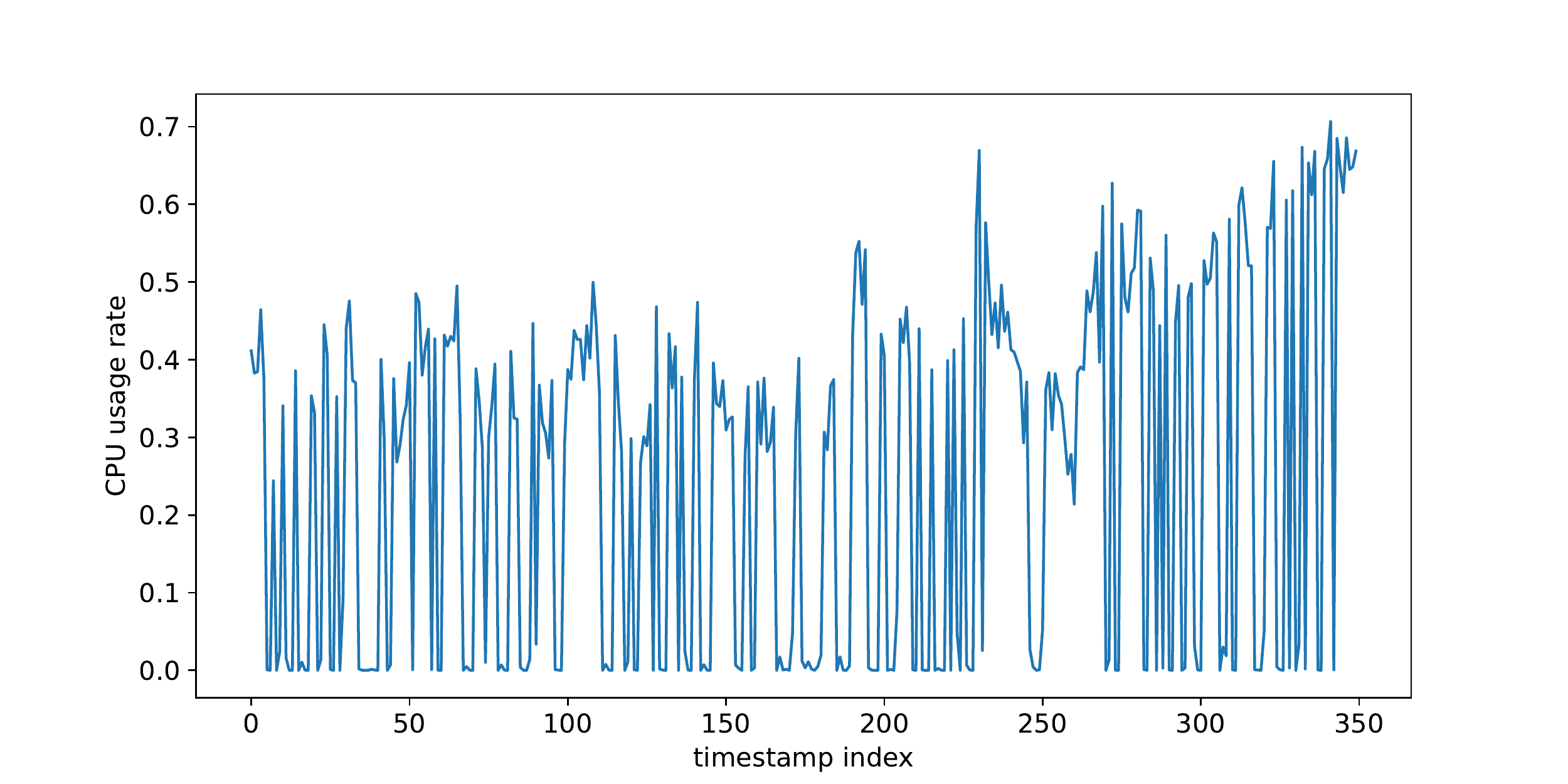}
    \caption{The first 350 CPU usages in the dataset}
    \label{fig:cpu-traces}
\end{figure}

\begin{table}[ht] 
\caption{Overview of the Google cluster traces dataset}
\centering
\renewcommand{\arraystretch}{1.5}
\begin{tabular}{|c|c|c|c|c|c|c|}
\hline\hline
Metric & Result\\ [0.2ex] 
\hline   
Unique jobs count & 672,074\\
Unique machines count & 12,583\\
Average CPU usage rate & 0.0205\\
Average canonical memory usage & 0.0201\\
Average disk I/O time & 0.002\\
Average local disk space used & 8.982e-05\\
\hline\hline  
\end{tabular} 
\label{tab:google-traces-metrics} 
\end{table} 
As discussed earlier, Google traces dataset provides resource usage information for each task over time. On the other hand, our method is centered on predicting machine resource usage; to create the dataset needed for this purpose, we aggregated task usage in every period, calculating the total machine resource usage in the period. Figure \ref{fig:cpu-traces} depicts the first 350 total aggregated CPU usages over time for the chosen machine. 18745 CPU usages are available in the dataset.

To prepare the data for prediction, we analyzed several factors: First, we examined the features for missing values. No missing values were found in CPU usage rate, canonical memory usage, and local disk space; however, disk I/O time included missing values, so we replaced these values with a linear interpolation using existing data points. Second, feature values are normalized between a range of zero and one to overcome model fitting problems. We utilized a MinMax scaler for normalizing, defined in Equation \ref{eq:minmax}, where $x$ is the feature being scaled.
\begin{equation}
    x_{scaled} = \frac{x - x_{min}}{x_{max} - x_{min}}
\label{eq:minmax}
\end{equation}
\subsection{Evaluation Metrics}
We used mean absolute error (MAE) and root mean squared error (RMSE) to evaluate our results and compare our model with other models. These errors are defined as:
\begin{equation}
    MAE = \frac{1}{n} \sum_{i=1}^{n} |y_i - y'_i|
\end{equation}
\begin{equation}
    RMSE = \sqrt{\frac{1}{n}\sum_{i=1}^{n}(y_i - y'_i)^2}
\end{equation}
where $y_i$ and $y'_i$ show the actual and the predicted output, respectively.
\subsection{GRU Training and Prediction}
\begin{figure}[hbt]
    \centering
    \includegraphics[width=0.47\textwidth]{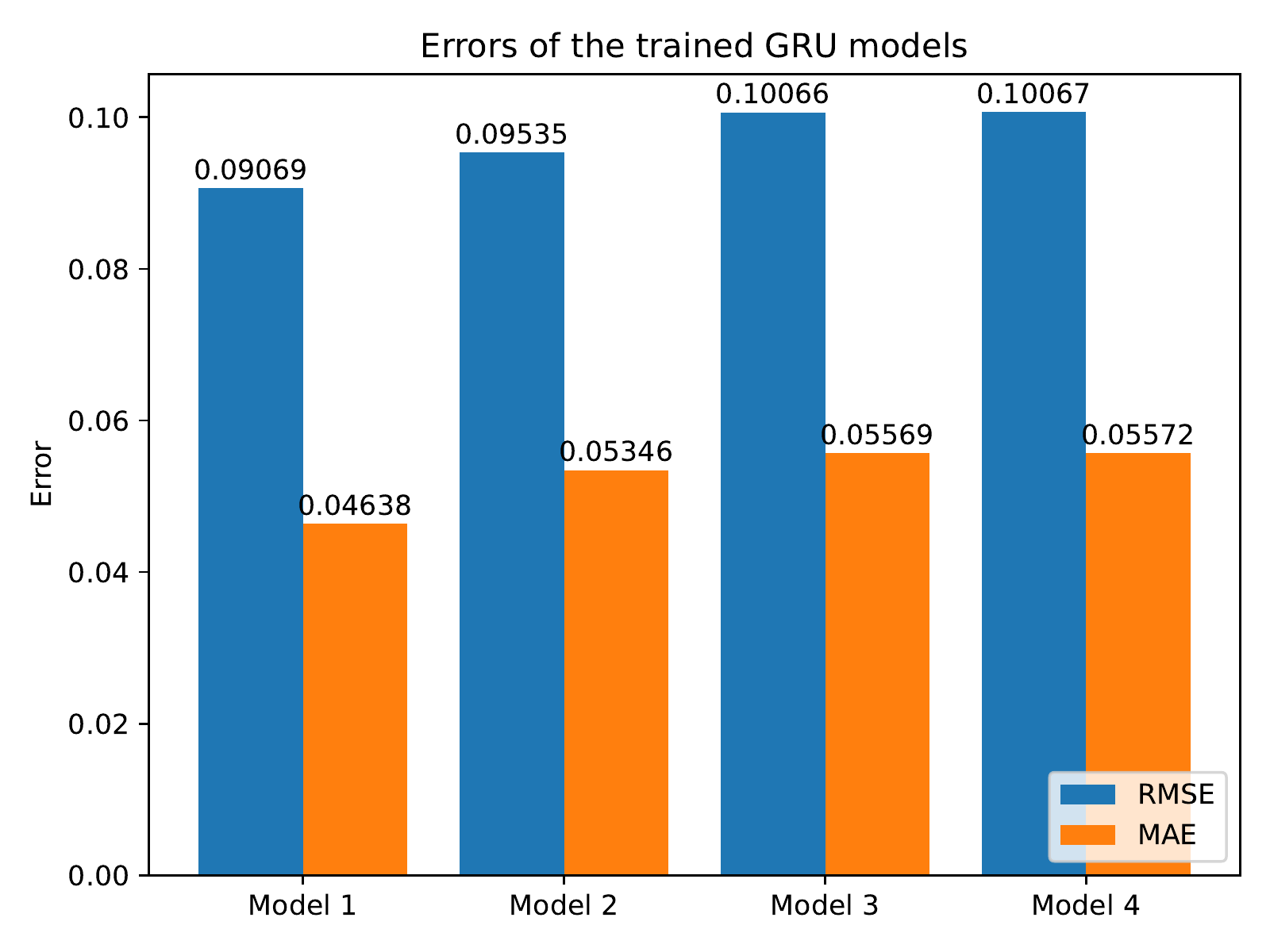}
    \caption{The best GRU models trained with different hyperparameters}
    \label{fig:gru-best-models}
\end{figure}

\begin{figure}[hbt]
    \centering
    \includegraphics[width=0.47\textwidth]{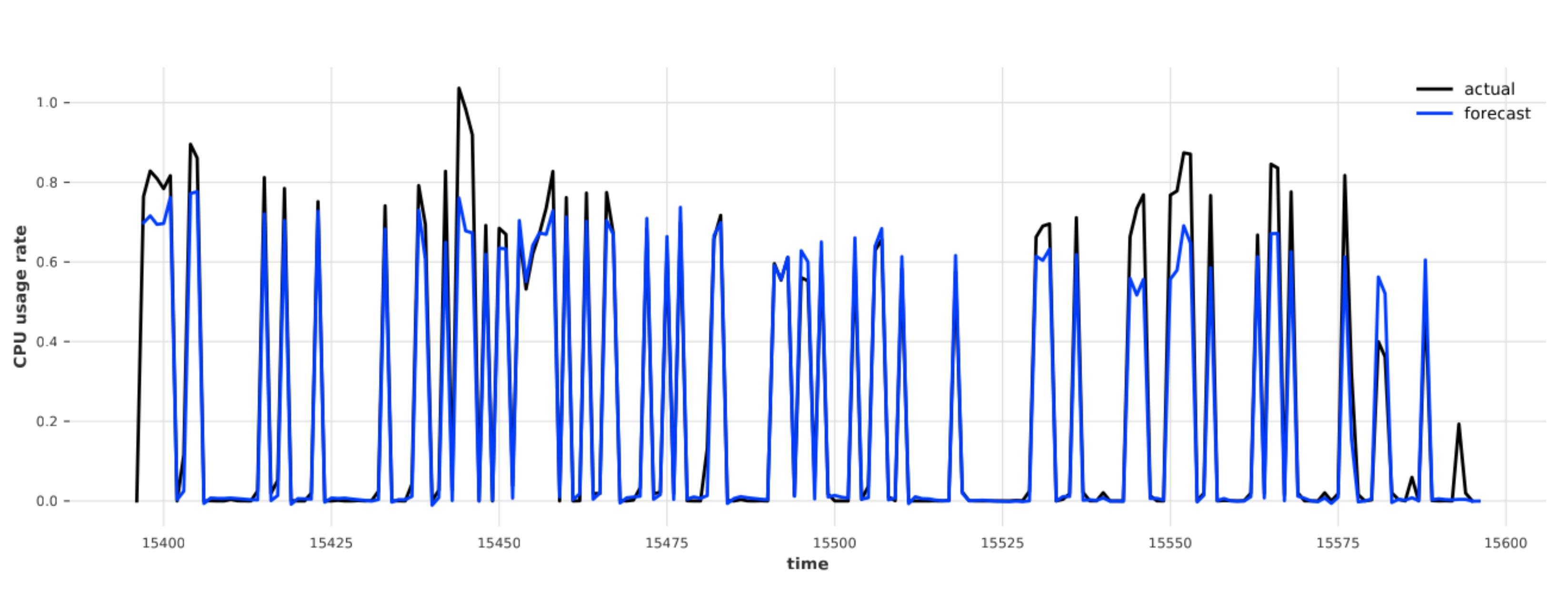}
    \caption{The CPU usage rate forecasts using the trained GRU model with 64 cells in each hidden layer, 3 GRU layers, and a previous step of 12}
    \label{fig:gru-model1-forecasts}
\end{figure}

We used the Darts \cite{JMLR:v23:21-1177}, PyTorch Lightning \cite{falcon2020pytorchlightning}, and PyTorch \cite{NEURIPS2019_9015} frameworks to train our forecasting model. To develop the GRU model, we split the complete workloads data into training subsets with 80\% and 20\% of the data. Next, we used the GridSearch algorithm to find the best hyperparameters for the model. GridSearch finds the most accurate model by comparing different combinations of the passed parameters and the respective error rates. The size of the GRU's hidden layer neurons, the number of RNN layers, and the number of past steps fed into the model were analyzed with GridSearch to develop the most accurate model. Subsequently, we analyzed 18 models with different values for the mentioned metrics. The size of the hidden layer neurons is selected from the set {32, 64}, the number of RNN layers is chosen from the set {1, 3, 5}, and we used {4, 8, 12} as the number of past steps used in the training process. Figure \ref{fig:gru-best-models} depicts the error rates of the four best models after the GridSearch process. Moreover, Table \ref{tab:best-gru-models} shows the hyperparameters chosen for these models. Model~1, which has 3 RNN layers with 64 hidden neurons in each layer, outperforms the other models in RMSE and MAE error rates. Model~2, having 1 RNN layer with 32 hidden neurons and a previous step size of 12, similar to Model~1, is the next best model, according to the results. Model~3 and Model~4 are the next most accurate models and have near RMSE and MAE error rates. All these models use a three-step prediction, resulting in a 15-minute early forecast due to the 5-minute gap between data points in the Google cluster traces dataset.
\begin{table}[ht] 
\caption{The best GRU models' hyperparameters}
\centering
\renewcommand{\arraystretch}{1.5}
\begin{tabular}{|c|c|c|c|c|c|c|}
\hline\hline
Model & Hidden layer size & GRU layers & Previous steps\\ [0.2ex] 
\hline   
Model 1 & 64 & 3 & 12\\
Model 2 & 32 & 1 & 12\\
Model 3 & 32 & 1 & 4\\
Model 4 & 32 & 1 & 8\\
\hline\hline  
\end{tabular} 
\label{tab:best-gru-models} 
\end{table} 
Due to the error rates, we used Model~1, the best RMSE and MAE model, as our base GRU model in our research. In addition, Figure \ref{fig:gru-model1-forecasts} shows the forecasted workloads and the actual observations plot. According to the forecasts, the true CPU usage rate observations are accurately predicted using the model trained with our method.

\begin{figure}[hbt]
    \centering
    \includegraphics[width=0.47\textwidth]{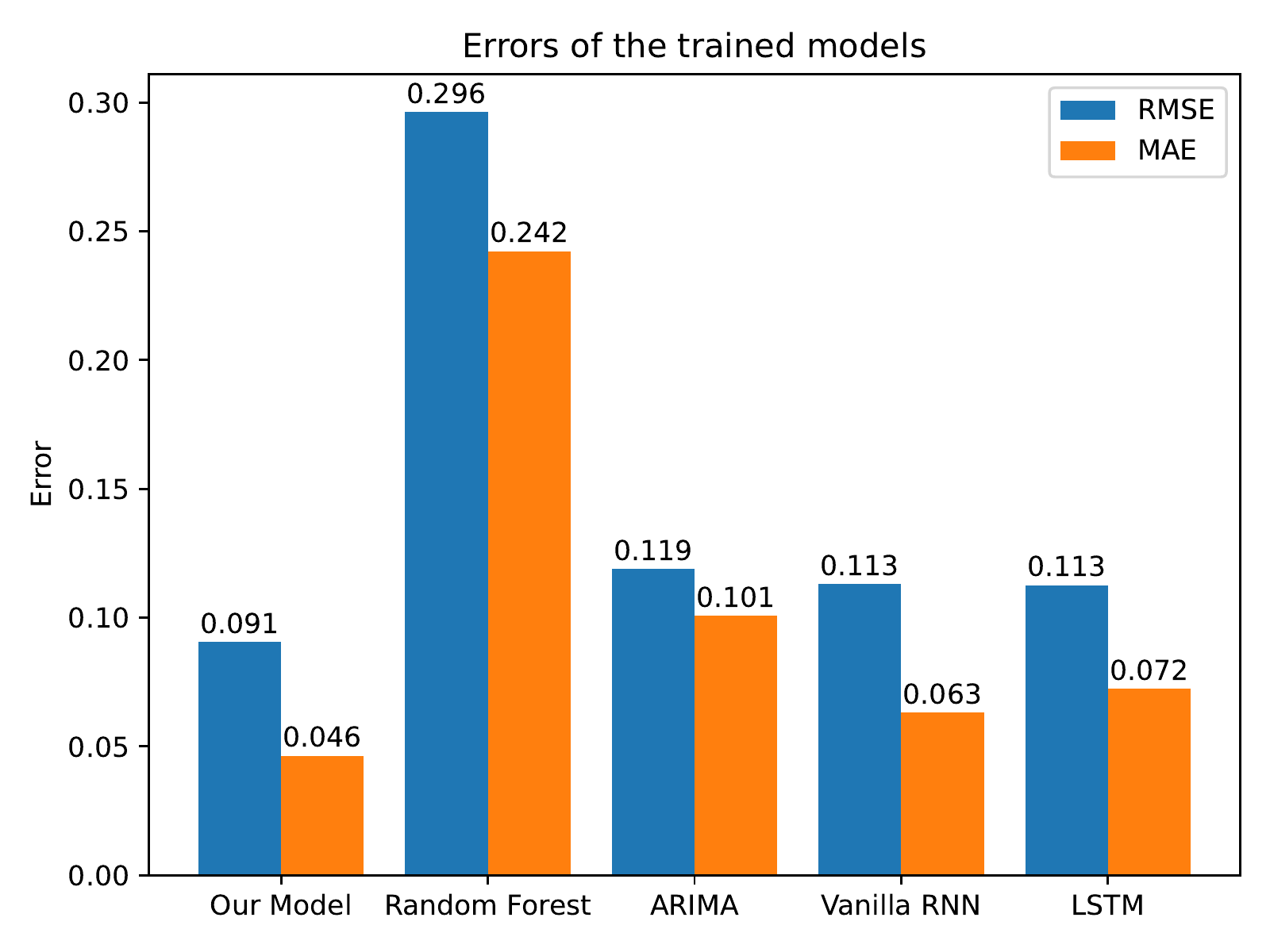}
    \caption{The RMSE and MAE errors of our model compared with the baseline models}
    \label{fig:models-compared}
\end{figure}

To compare the GRU model's accuracy, we trained other models using different methods, such as random forest, ARIMA, vanilla RNN, and LSTM; these models' RMSE and MAE error rates are provided in Figure \ref{fig:models-compared}. According to the results, our model outperforms the other methods in MAE and RMSE rates. The random forest model has the highest error rate in the results and performs worse than other models. Moreover, the ARIMA model has a higher error rate compared with the neural network-based models such as vanilla RNN and LSTM. The vanilla RNN and LSTM models performed similarly regarding RMSE error rates due to their network structure similarity. However, the RNN model had fewer MAE errors than the LSTM model, which we think is because of the different hyperparameters when training.
\subsection{Pruning Results}
We analyze different pruning approaches to increase the model speed with a slight decay in accuracy. We compared two main methods: L1 norm structured and random structured pruning. We started with the initial model trained with the hyperparameters chosen from the GRU training process, and we increased the pruning amount to observe the effect on the RMSE and MAE error rates. In addition, we analyzed the impact of the pruning amount on the prediction speed.

\begin{figure*}[hbpt!]
    \centering
    \includegraphics[width=\textwidth]{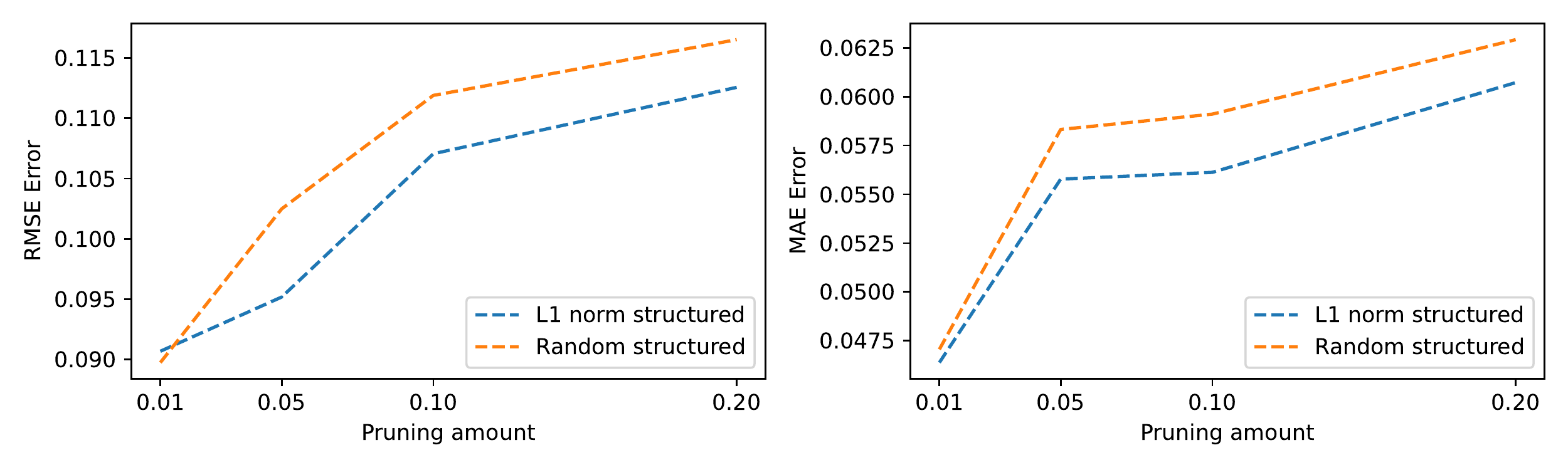}
    \caption{The RMSE and MAE error rate of L1 norm and random structured pruning methods compared after pruning amount changes}
    \label{fig:pruning-acc}
\end{figure*}

Figure \ref{fig:pruning-acc} depicts the MAE and RMSE error rates for different pruning amounts and methods. As seen in the results, the L1 norm structured pruning is performing better regarding MAE and RMSE error rates. Observing the RMSE error rate plot, we find that the 0.05 pruning amount did not affect the RMSE error rate of the L1 norm much, and the RMSE rate is very similar to the initial unpruned model. However, the increase has a more negative effect when the pruning is done randomly. Moreover, The plot shows that the rise of pruning amount to 0.1 and 0.2 has a significant error rate increase for both models. On the other hand, speaking about the MAE errors plot, we do not see a considerable MAE error rate increase after raising the pruning amount to 0.1 since the error is relatively the same for the 0.05 and 0.1 pruning ratios. Similar to the RMSE results, increasing the pruning amount to 0.2 impacts the MAE error considerably for both L1 norm and random methods.
\begin{figure}[H]
    \centering
    \includegraphics[width=0.47\textwidth]{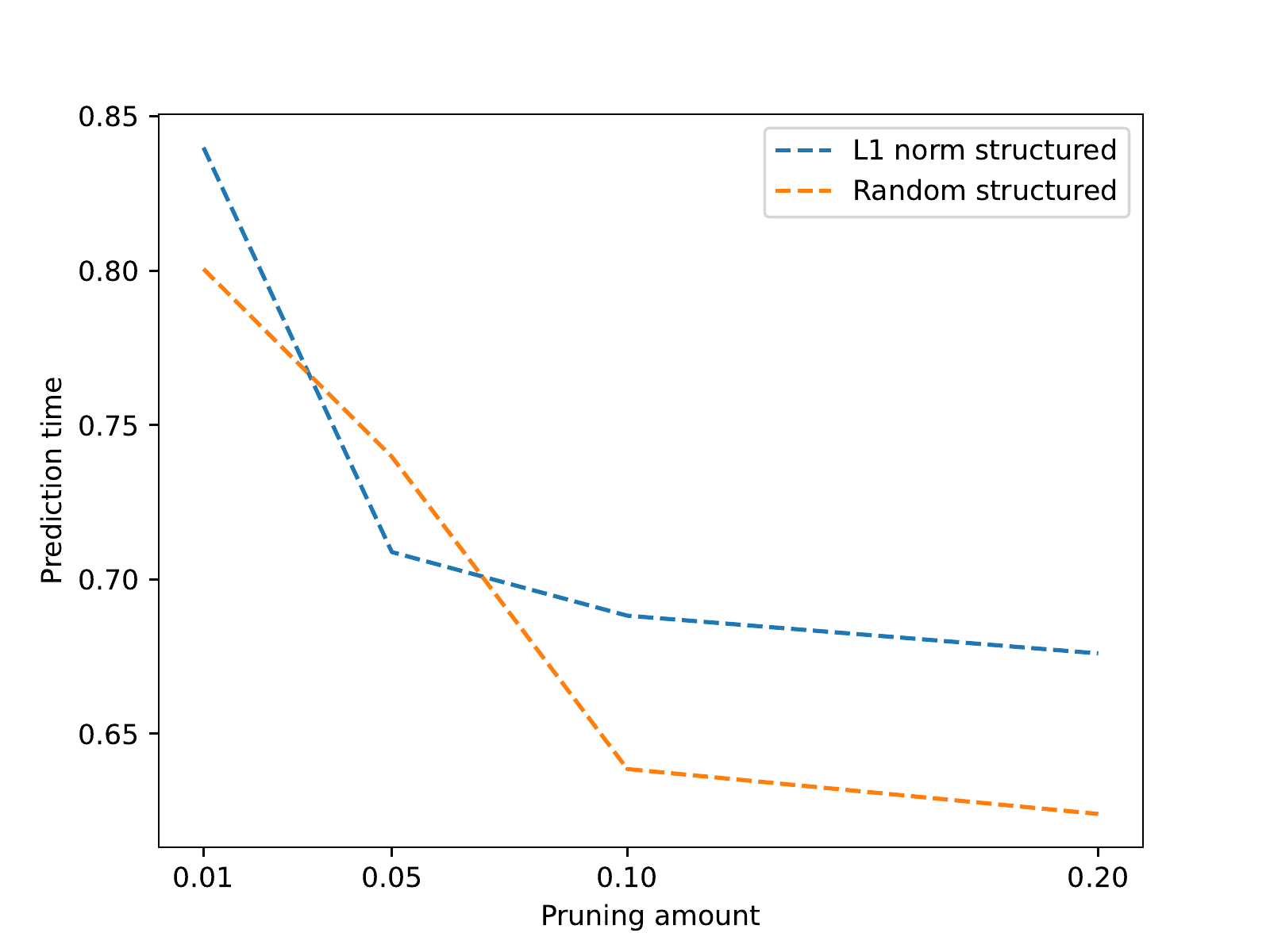}
    \caption{The prediction time results using L1 norm and random structured pruning methods}
    \label{fig:pruning-speed}
\end{figure}
Figure \ref{fig:pruning-speed} shows the prediction speed of both methods with different pruning amounts. According to the prediction time results, a lower prediction time is obtained by increasing the amount of pruning. However, the L1 norm and random methods differ regarding prediction speed; the initial L1 norm pruning method has a higher prediction time compared with the random pruning method. Increasing the pruning amount to 0.05, we observe that the L1 norm prediction time notably decreases, even lower than the time the random pruning method takes at the 0.05 pruning point. By raising the pruning amount more, the random method significantly decreases the prediction time, which is not seen in the L1 norm results; The L1 norm method's prediction speed remains relatively the same after increasing the pruning amount past the 0.05 value.

Considering the error rate results and the speed-up measurements, we see that the L1 norm method has better results regarding accuracy; however, it falls back from the random method in terms of speed in higher pruning levels. We find that a pruning amount of 0.05 for the L1 norm method is the best choice since it has a good RMSE and MAE error rate, and no increased speed-up is seen after this value according to the prediction time results. In addition, the random structured pruning works better using a pruning amount of 0.1, which increases the model speed significantly. Generally, the choice of pruning amount is a trade-off between the model's accuracy decay and speed-up gains; Therefore, one must consider the requirements to choose the appropriate pruning method and amount.
\subsection{Online Learning Results}
\begin{figure*}[hbt]
    \centering
    \includegraphics[width=\textwidth]{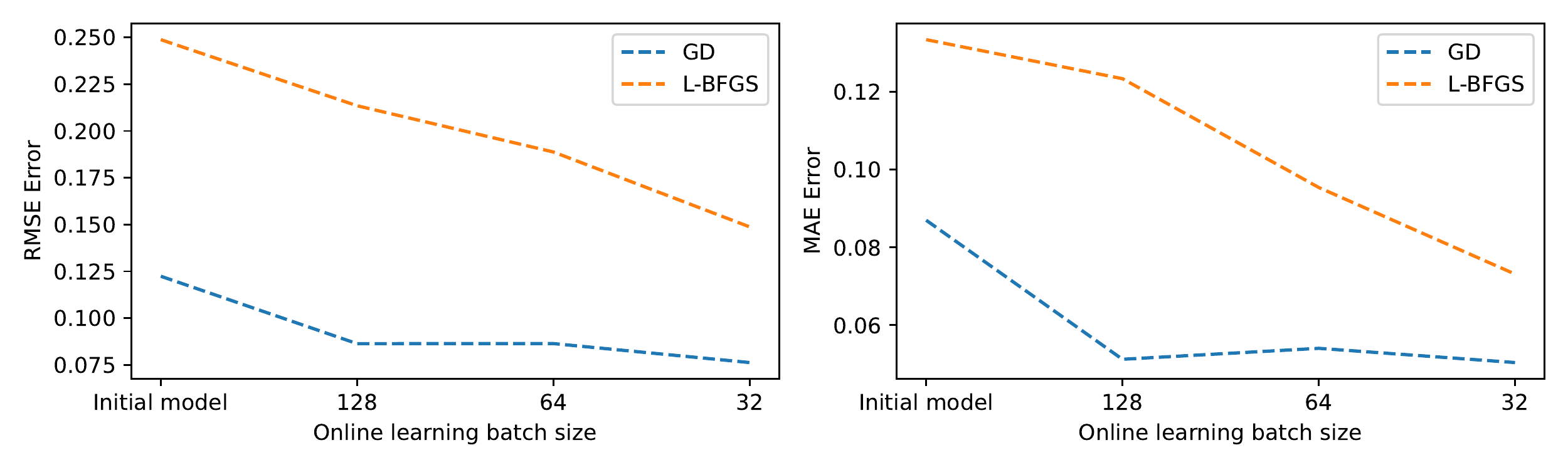}
    \caption{The RMSE and MAE error rates of GD and L-BFGS optimization methods when online learning batch size is changed}
    \label{fig:online-acc}
\end{figure*}

This section describes our online learning method's results. We used GD and L-BFGS optimization mechanisms to analyze different optimization approaches for adapting the model. Moreover, the online learning batch size's effect on the RMSE and MAE errors is analyzed. The online learning batch size is the number of observations the model waits to be available to perform online adaptation. Figure \ref{fig:online-acc} depicts the results of the two optimization approaches with different online learning batch sizes.
Taking into account the RMSE error results, we find that both model accuracies improve after performing online adaptation compared with the initial model with no online learning method; This is because online learning considers the new patterns available in the observations, which the initial model did not. The GD method performed better regarding error rates. Moreover, the L-BFGS approach took more time to converge and performed poorly. Furthermore, the model error rates improve when the online learning batch size decreases, enabling the model to learn new patterns faster and utilize them on the following observed workloads. Both models show the same pattern of decreases in errors with smaller batch sizes; however, the GD method's error rate does not change much when we decrease the batch size from 128 to 64.

We see a similar pattern in the MAE results. GD performs better regarding the MAE error rate. Like the RMSE results, both models' error rates improve by decreasing the online learning batch sizes, and the online learning models outperform the initial model with no online learning mechanism. Again, the GD model's MAE error does not notably change by decreasing the batch size from 128 to 64. Even though the L-BFGS error rates are worse compared with the GD method, the decrease in the error rate is faster when we decrease the batch size. A smaller batch size is significantly better for the L-BFGS method and helps it perform better.

\begin{figure}[H]
    \centering
    \includegraphics[width=0.47\textwidth]{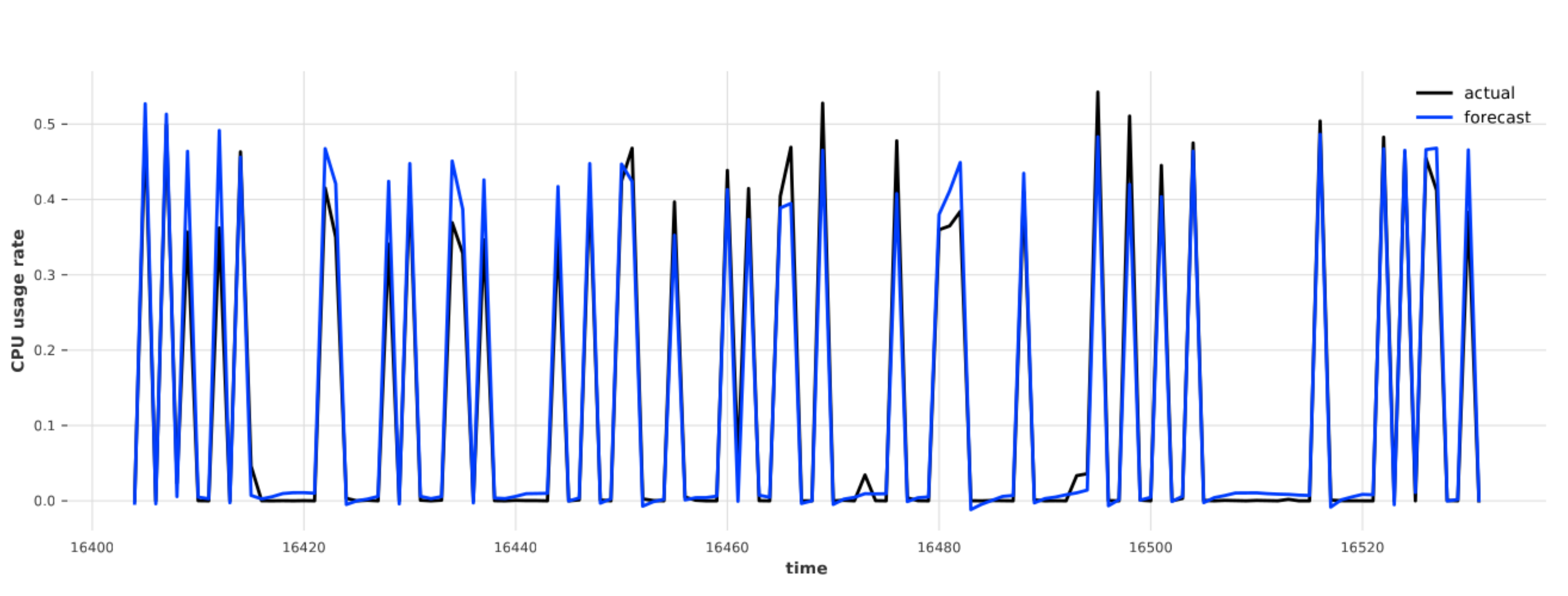}
    \caption{The plot of CPU usage forecasts for batch number 10 after online adaptation using a batch size of 128 observations}
    \label{fig:online-preds}
\end{figure}

In addition, Figure \ref{fig:online-preds} shows the plot of forecasts and actual observations for batch number 10 when online adaptation is made using a batch size of 128 observations. As we can see, the forecasts after online prediction match the actual trends in the workloads and predict them well.
\section{Conclusion}
\label{conclusion}
Our research developed a multivariate multistep GRU neural network to forecast workloads accurately. First, the best performing GRU model is created on the Google traces dataset using GridSearch in the hyperparameters space. We find that a GRU model with 3 GRU layers, 64 neurons, and 12 observations performs best. Next, we compared the trained GRU model with several baseline models. We observe that the trained GRU model outperforms the other baseline models. Furthermore, we used two pruning methods: L1 norm structured and random structured, to create a sparser model to predict workloads faster. We notice that a higher pruning amount results in slight error rate increases and better prediction times, which is expected. We note that the L1 norm structured method performs better regarding prediction accuracies, and the random method predicts faster in higher pruning amounts. Finally, we analyzed the online adaptation of the created model using two optimization techniques: GD and L-BFGS. After changing the online learning batch size, we compared the RMSE and MAE results. The results show better performances after decreasing the batch size. Also, we observe that GD performs better in both RMSE and MAE error rates. In conclusion, our approach enhances the accuracy and speed of prediction using a multivariate multistep GRU model in addition to pruning and online adaptation.

For future work, the following can be done:
\begin{itemize}
    \item Analysis of the effect of the prediction horizon on the workload forecast accuracy and scheduling decisions
    \item Analyses of the pruning speed achievements in a real-time cloud system
    \item The comparison of online adaptation frameworks in a real-time workload system regarding accuracy enhancement
\end{itemize}
\bibliographystyle{IEEEtran}
\bibliography{IEEEabrv,article}

\end{document}